# INTEGRATION OF MICRO-ELECTRO-MECHANICAL DEFORMABLE MIRRORS IN DOPED FIBER AMPLIFIERS

*D. Bouyge, D. Sabourdy, A. Crunteanu, P. Blondy, V. Couderc, J. Lhermite, L. Grossard and A Barthélemy*

XLIM Research Institute, University of Limoges, 123 Av. Albert Thomas, 87060 Limoges, France

**ABSTRACT**

We present a simple technique to produce active Q-switching in various types of fiber amplifiers by active integration of an electrostatic actuated deformable metallic micro-mirror. The optical MEMS (MOEMS) device acts as one of the laser cavity reflectors and, at the same time, as switching/ modulator element. We aim to obtain laser systems emitting short, high-power pulses and having variable repetition rate. The electro-mechanical behavior of membrane (bridge-type) was simulated by using electrostatic and modal 3D finite element analysis (FEA). The results of the simulations fit well with the experimental mechanical, electrical and thermal measurements of the components. In order to decrease the sensitiveness to fiber-mirror alignment we are developing novel optical devices based on stressed-metal cantilever-type geometry that allow deflections up to 50 μm with increased reflectivity discrimination during actuation.

## 1. INTRODUCTION

For pulse generation in a conventional Q-switched fiber laser, a passive or active modulator (saturable absorbers, or acousto-optic, electro-optic or mechanical components) has to be introduced into the cavity [1]. Although these conventional solutions are based on solid, mature technologies, most of them present inherent disadvantages that restrain their integration in miniature, compact laser systems: degradation of the beam quality, high insertion losses for the acousto-optic modulators [2], high voltages and low modulation frequencies for the electro-optics solutions [3], bulkiness for mechanical choppers, low laser power levels operation for piezoelectric Bragg gratings systems [4] or lack of control of frequency and pulse width for the passive modulators. We previously reported [5] on active Q-switching demonstration for an erbium-doped fiber laser (EDFA) using a deformable metallic micro-mirror like the one represented in Figure 1. As explained there, the micro-mirror consists of a 500 nm-thick gold membrane (bridge) suspended over an actuation electrode placed 2.2 μm underneath the membrane and covered with a dielectric thin film ($Al_2O_3$, 200 nm thickness) for electrical isolation during actuation

The laser system incorporated the optical MEMS device operates at frequencies between 20 and 120 kHz and generate short pulses (with FWHM down to 326 ns) and high peak powers, up to 100 times greater than the continuous emission [5].

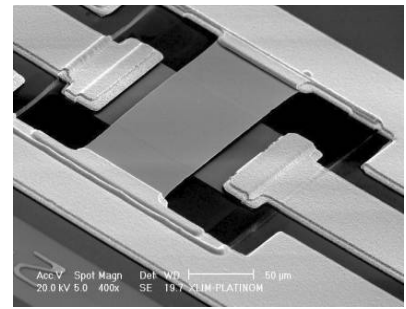

**Figure 1.** Scanning electron micrograph (SEM) of a typical metallic micro- mirror

The MOEMS device is small, compact, highly reflective and achromatic, allowing a great integration potential. The simplicity of this pulse generation technique makes it suitable for being implemented in more complex set-ups including solid-state micro-lasers, multi-wavelength fiber lasers or different families of fiber lasers (Yb- or Er-Yb-doped) actuated independently or synchronous for wavelength mixing/ tuning applications.

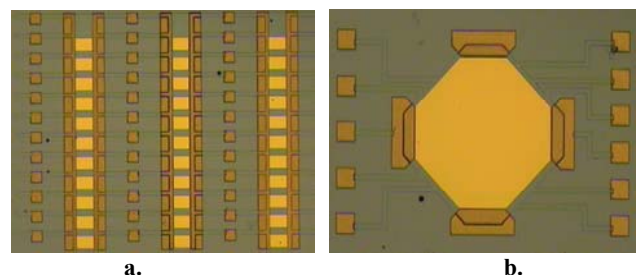

a.  b.
**Figure 2.** a) Matrix of micro-mirrors actuated independently or simultaneously and b) a 1.5x1.5 mm² large-area membrane

The images presented in Figure 2 shows several designs based on this bridge-type membrane. Figure 2a is a matrix of 44 micro-mirrors, which can be integrated with several parallel fiber amplifiers for generation of synchronized pulse trains. The larger membrane (1.5x1.5 mm$^2$), which can be actuated by independent electrodes, (Figure 2b) is planned to be used with larger, multimode laser beams for mode selection.

Here we report on the thermo-mechanical behaviour of the membrane mirrors through FEA simulations and experimental measurements. We present recent results of different fiber laser systems coupled with the MOEMS devices and the design and fabrication of devices with improved capabilities developed to optimise the laser Q-switch generation.

## 2. BRIDGE-TYPE MEMBRANE

### 2.1. Mechanical measurement and FEM simulation

The most critical parameter for generating high-energy pulses in a Q-switched fiber laser is the modulator speed. In our case this parameter is determined, in a first-order approximation, by the mirror's mechanical primary resonant frequency (directly linked to the switching time of the mirror). In-plane tensile stresses, arising from device micro fabrication and temperature changes, may result in a significant increase of the resonant frequency and of the actuation voltage [6]. In order to determine the effects of the build-in stress and operating temperature on the mechanical resonant frequency, we simulated the behavior of the membranes having different dimensions and under different temperatures, using MultiPhysics package of ANSYS. The experimental resonant frequency values were recorded by using an electro-mechanical method presented in [7]. The material properties (considered to be isotropic in the XYZ directions) of the membranes are listed in Table I. Figure 3 shows the simulation results of the mechanical resonant frequency of gold- and aluminum-type membranes (0.5 µm in thickness and 80 µm in width) versus their length at room temperature (RT). We noticed that for widths values in the range of 50 to 160 µm, this parameter is not influencing the resonant frequency (variation of less than 3%). In the case of the gold layers the experimental measurements are very well fitted when we applied a value of 30 MPa uniaxial tensile stress to the membranes.

**Table I.** Material properties used for the simulation of metallic membrane thermo-mechanical behavior

| Material property | Au | Al |
|---|---|---|
| Young's modulus (GPa) | 78 | 70 |
| Thermal expansion coefficient ($10^{-6}$ K$^{-1}$) | 14,2 | 23,5 |
| Poisson ratio | 0,42 | 0,345 |
| Material density ($10^{-18}$) | 19300 | 2700 |

The gold- type mirrors have a relatively high mechanical resonant frequency at RT, ranging from ~ 65 kHz (for the 240 x 160 µm$^2$-area membranes) up to 170 KHz (for the 120 x 60 µm$^2$ ones).

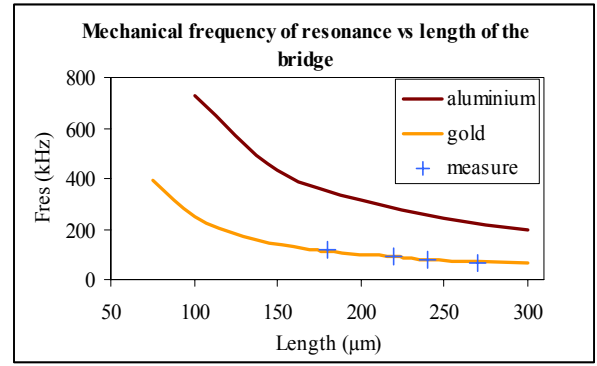

**Figure 3.** Mechanical primary resonant frequency RT as a function of the membrane length (points for gold- type bridge experimental measurement and for gold- and aluminum-type bridges from FEA simulation)

We are currently designing and implementing similar types of optical switching elements, which are intended to be faster. Lowering of the switching time of the mirrors will lead to narrower laser pulses with higher peak powers as well as higher pulse repletion rates. We produced smaller bridges with lengths ranging from 75 µm up to 150 µm, and we are currently developing aluminum-type bridges using the same fabrication process. According to the curves on Figure 3 the mechanical resonant frequency at RT of the smaller membranes will take values between 150 kHz and 400 kHz for the gold-type devices while for the aluminum-type ones the frequencies will be higher than 400 kHz. Figure 4 shows the temperature evolution of the resonant frequency for a 160 x 220 µm$^2$ area gold mirror undergoing a cooling-heating cycle between RT and liquid- nitrogen (LN$_2$) temperature (77 K). During the cooling cycle the resonant frequency is increasing up to 250 KHz but the membrane recovers its mechanical behavior when heating back to RT.

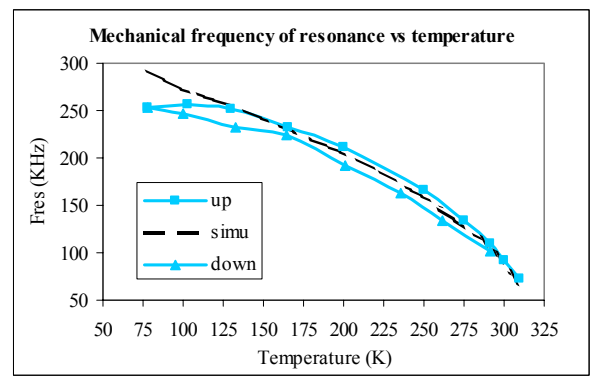

**Figure 4.** Experimental measurement of F$_{res}$ with temperature between RT and 77 K and FEA simulation (30 Mpa tensile stress) for the 160 x 220 µm$^2$ membrane

Low temperatures increase the tensile stress of the membrane, which explain higher $F_{res}$ values. We observed that at higher temperatures (up to 370 K) the stress turn into a compressive one that leads to buckling of the bridge and device failure.

**2.2. Fiber laser systems coupled with the optical MEMS**

*2.2.1. Opto-mechanical characterizations of the micro-mirrors*

The laser beam focused on the membranes is deflected by ~9° during switch actuation, with a good reflectivity discrimination between the on- and off- actuation states, which is enough for loss modulation of the laser cavity. The laser power handling of the micro-mirrors was investigated by placing them in an Er/ Yb co-doped fiber laser. The amplifier was side- pumped by a laser diode emitting at 915 nm with powers up to 5 W. The experiments ware carried out for various levels of emitted laser power. In the non-actuated, off- state, the membranes seem to sustain quite well laser powers up to 1 W (the maximum power delivered by the laser) in CW regime. In the Q-switch regime, for mean powers higher than 900 mW, the membrane degrades leading to a brutal fall of the power.

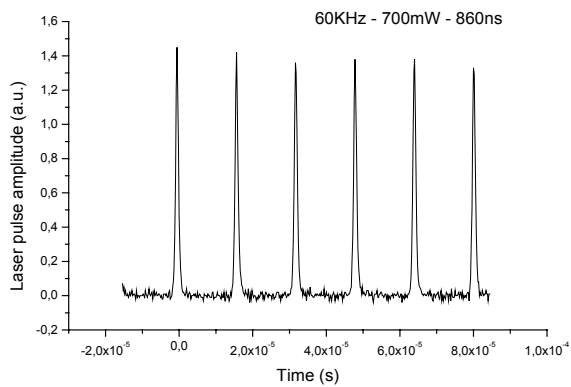

**Figure 5.** Pulse train generation using an 160 x 220 µm² MOEMS at 60 kHz actuation frequency

In Figure 5 is presented the generation of laser pulses of 860 ns FWHM at 60 kHz actuation frequency and 700 mW mean laser power, implying pulse peak powers of ~14 W.

*2.2.2. Dual-wavelength fiber amplifier*

Because of the achromaticity of the membrane, the MOEMS element can be used with different types of amplifiers emitting at different wavelengths. The dual-wavelength fiber laser system depicted in Figure 6 consists in two synchronized Q-switched fiber lasers (an EDFA and an Yb- doped fiber amplifier). Both lasers are core pumped through a wavelength division multiplexer (WDM) by a laser diode emitting up to 100mW at 980nm. At the ends of the two amplifiers, the two signals are mixed thanks to a WDM (1060/1550). The output coupler of the laser system has a 4% reflectivity (right angle cleaving of the end of the WDM fiber). The dual laser system is closed at the other end by the optical MEMS element (as back achromatic cavity mirror and Q-switch modulator).

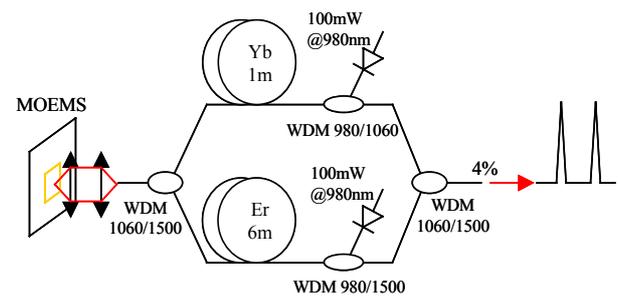

**Figure 6.** Set-up of the laser fiber system emitting dual-wavelength pulses

Figure 7 presents the generation, using this system, of perfectly superposed pulse trains coming from both lasers having FWHM of 820 ns at an actuation frequency of 30 kHz. Figure 8 show the spectral components of each pulse peaks and one can notice the well-known emissions of Yb and Er at ~1060 nm and at ~1550 nm, respectively. We obtained, thus, a laser system emitting synchronized pulse trains with two components in the spectral domain.

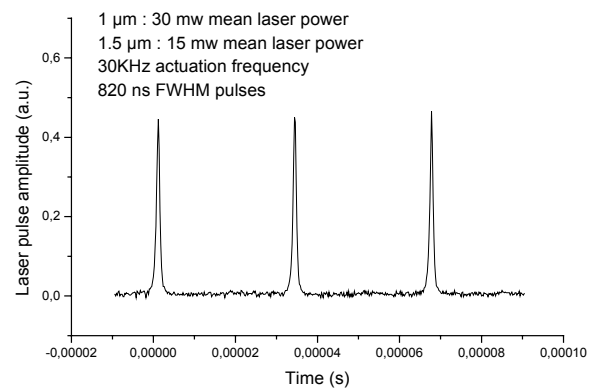

**Figure 7**. Pulse train generation of the dual-wavelength fiber laser system

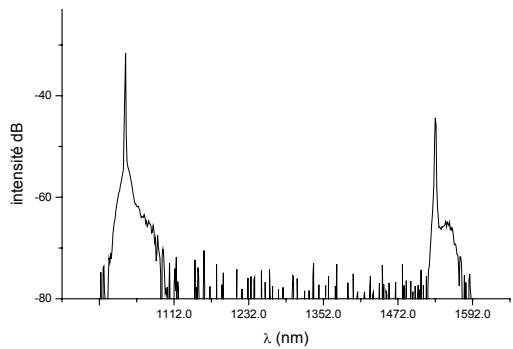

**Figure 8.** Spectral emission of pulses for the Q-switched dual-laser system

*2.2.3. Active Q-switching in an erbium-doped fiber laser (EDFA)*

In the architecture of the EDFA presented in [5] the back end emission of the laser is coupled to the MOEMS through an imaging system based on two confocal micro-lenses. In this way we obtained pulses with peak powers of several watts and repetition rates that can be continuously tuned from 20 kHz to 120 kHz. The duration (FWHM) of the generated pulses varies from 326 ns up to 1 µs.

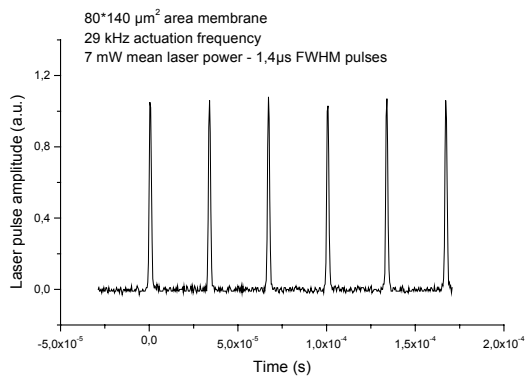

**Figure 9.** Pulse train generation without an imaging system

Taking into account the low dimensions of the MOEMS elements, it will be very interested to realize an even more compact device, with higher integrability. For that, it is necessary to be freed from the imaging system. We observed that without the imaging system, the re-injection of the laser beam in the fiber amplifier is somehow reduced, resulting in a significantly decrease of the mean laser power. In spite of this reduction the beam modulation is still fast enough for the laser to pass in the Q-switching regime even in this simple configuration.

Using the same micro-mirror (80 x 140 µm² area membrane) as presented in [5] the EDFA emits 1.4 µs FWHM pulses, at 39 kHz actuation frequency with a 7 mW mean power laser, which represents pulses with peak power of only 150 mW (Figure 9).

### 3. CANTILEVER-TYPE MICRO-MIRRORS

For obtaining higher reflectivity discrimination during MOEMS actuation we designed and fabricated micro-mirrors based on a cantilever-type design (Figure 10). The design is meant to avoid the use of an additional imaging system leading to smaller, integrated packaged system. The principle of operation of this MEMS structure has been presented in [8] and next developed in [9]. It is based on a curled metallic membrane above a conductive electrode. The stressed, up-warded, membrane is fabricated on a low resistivity Si substrate (acting as lower electrode) covered with 1 µm thermally-grown $SiO_2$ layer (as dielectric isolator). When applying a tension between the substrate and the cantilever, this one is attracted toward substrate due to the generated electrostatic force.

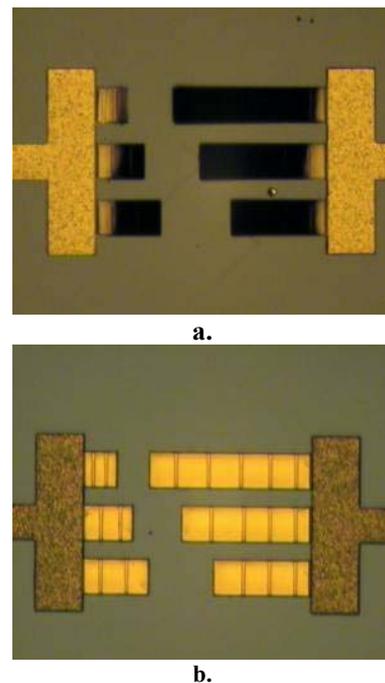

**Figure 10.** Cantilever-type mirrors in a) OFF-state and b) ON-state

The fabrication process started by the deposition of a two steps sacrificial layer that allows to separate the metal layer 0.6 µm apart from the substrate, and make 1.5 µm deep corrugations in order to provide longitudinal stiffness to the structure and successful control of the initial deformation, and to decrease the contact area between the membrane and the $SiO_2$ dielectric during actuation. The metallization is done using a Cr / Au (40 Å / 800 Å) seed layer, gold-electroplated up to 1.5 µm. Next a 100 Å Cr stress layer is deposited and patterned on the foldable areas, in order to provide an appropriate stress gradient in the foldable areas. Finally, the component is released and the MOEMS is dried in a critical point drying system.

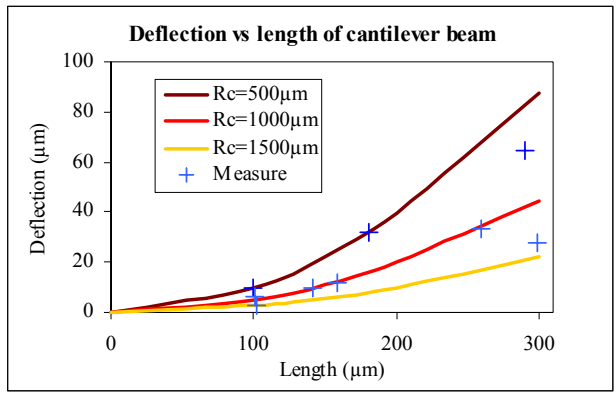

**Figure 11.** Deflection versus length curves for cantilever-type MOEMS (experimental measurement and analytical solution with various radius of curvature)

We fabricated cantilever-type devices with different dimensions (from 50 x 50 µm$^2$ to 300 x 500 µm$^2$) and different designs (rectangular and triangular). The simulations of the mechanical resonant frequency for the rectangular membranes gives values from ~ 6 kHz up to 236 KHz at RT, depending on the dimensions. The profile of the cantilevers, measured using optical interferometry shows curvature radius between 500 µm and 1500 µm (Figure 11) depending on the design and on the components dimensions (Figure 12).

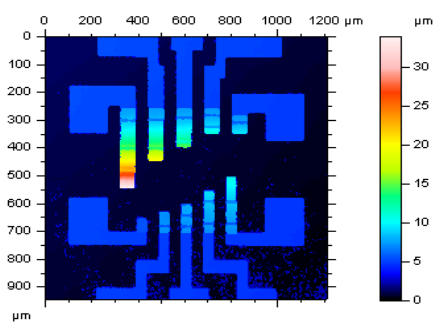

**Figure 12.** Top image of cantilevers-type profile

Figure 13 shows the 3D (a), the top image (b), and the profile (c) of an array of 200 µm in width and 250 µm in long triangular cantilever-type membranes. They have deflections up to 33 µm, which corresponds to a 1000 µm radius of curvature. The incident laser beam will be deviated by ~15° after reflexion on these particular micro-mirrors. The array can be used with laser beams having larger widths and may be a solution to improve the power handling capabilities of the MOEMS. The simulated mechanical resonant frequency of this particular structure is about 45 kHz.

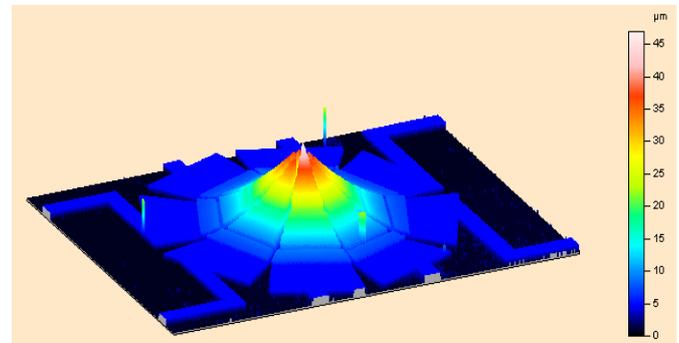

a.

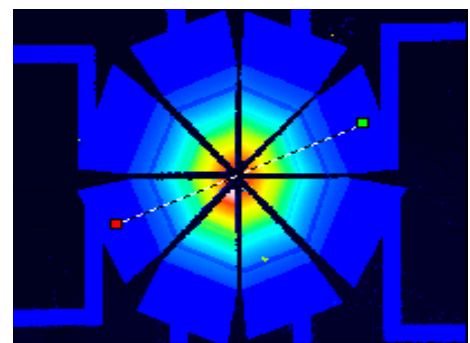

b.

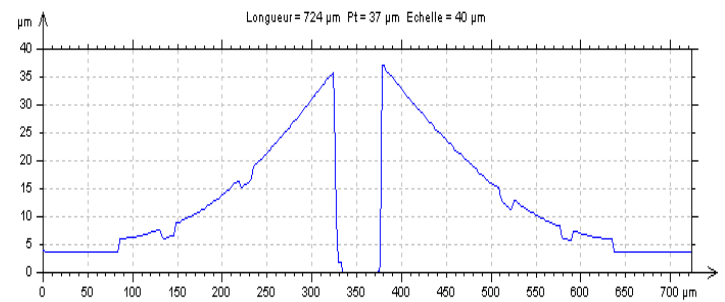

c.

**Figure 13.** 3D image (a), top image (b) and profile section obtained using optical interferometry

### 4. CONCLUSION

We developed micro-mirrors that can be integrated with various types of lasers amplifiers running at different wavelengths and with different architectures. A thermo-mechanical FEM model was developed to simulate the mechanical resonant frequency of the optical elements, and the results fit well with experimental measurement. Shortening of the switching time of the mirrors will lead to narrower laser pulses with higher peak powers as well as higher pulse repletion rates. Optimization of the laser Q-switch modulation using such micro-mirrors (bridge-type and especially, cantilever-type) will enable rapid development of applications like laser wavelength mixing, laser mode selection or multi-laser emission synchronization.

## 5. REFERENCES

[1] W. Barnes, ''Q-switched fiber lasers,'' in *Rare Earth Doped Fiber Lasers and Amplifiers*, New York, pp. 375–391, 1993.

[2] D D. Zalvidea, N.A. Russo, R. Duchowicz, M. Delgado-Pinar, A. Dýez, J.L. Cruz, M.V. Andres, *Optics Communications*, Volume 244, Issues 1-6, pp. 315-319, 2005.

[3] Ashraf F. El-Sherif and Terence A. King, *Optics Communications*, Volume 218, Issues 4-6, Pages 337-344, 2003.

[4] N. A. Russo, R. Duchowicz, J. Mora, J. L. Cruz and M. V. Andrés, *Optics Communications*, Volume 210, Issues 3-6, pp. 361-366, 2002.

[5] A. Crunteanu, D Bouyge, D Sabourdy, P Blondy, V Couderc, L Grossard, P H Pioger and A Barthélemy, "Deformable micro-electro-mechanical mirror integration in a fiber laser Q-switch system" *Journal of Optics A: Pure and Applied Optics*, accepted (2006)

[6] Yong Zhu, H.D.Espinosa, "Effect of temperature on capacitive RF MEMS switch performance-a coupled-field analysis", *Journal of micromechanics and mocroengineering*, pp. 1270-1279, 2004.

[7] D. Mercier, A. Pothier, P. Blondy "Monitoring mechanical characteristics of MEMS switches with a microwave test bench", *4th Round Table on Micro and Nano technologies for Space*, Noordwijk, Netherlands, 2003.

[8] S. Duffy, C. Bozler, S. Rabe, J. Knecht, L. Travis, P. Wyatt, C. Keast, M. Gouker, "MEMS microswitches for reconfigurable microwave circuitry," *IEEE Microwave and wireless components letters*, Vol 11, 3, pp.106-108, 2001.

[9] J. Muldavin, R. Boisvert, C. Bozler, S. Rabe, C. Keast, "Power handling and linearity of MEM capacitive series switches", *Microwave Symposium Digest*, IEEE MTT-S International, Volume 3, pp. 1915 – 1918, 2003.